\documentclass[times,astrobib,amssymb,usenatbib]{mn2e}
\usepackage{graphicx}
\usepackage{txfonts}
\usepackage{natbib}
\usepackage{supertabular}
\usepackage{longtable}
\usepackage{url}
\usepackage{dcolumn}
\usepackage{placeins}
\usepackage{multirow}
\usepackage{psfig}
\usepackage[utf8]{inputenc}

\newcommand{\be}{\begin{equation}}
\newcommand{\en}{\end{equation}}

\def\zabs{$z_{\rm abs}$}
\def\zem{$z_{\rm em}$~}

\def\mgi{Mg~{\sc i}~}
\def\mgii{Mg~{\sc ii}~}

\def\aliii{Al~{\sc iii}~}
\def\civ{C~{\sc iv}~}

\def\si2{Si~{\sc ii}~}
\def\feii{Fe~{\sc ii}~}

\def\kms{km~s$^{-1}$}
\title[Mg II absorption in  SDSS J133356.02+001229.1]{Dynamically evolving 
Mg~{\sc ii} broad absorption line flow in  SDSS J133356.02+001229.1}
\author[M. Vivek, R. Srianand, A. Mahabal, V. C. Kuriakose]{M. Vivek$^{1}$\thanks{E-mail:vivekm@iucaa.ernet.in},  R. Srianand$^{2}$,  A. Mahabal$^{3}$ \&  V. C. Kuriakose$^{1}$\\
$^{1}$Department of Physics, Cochin University of Science and Technology, Kochi 682022, India\\
$^{2}$Inter University Centre for Astronomy and Astrophysics, Pune 410007, India\\
$^{3}$Caltech, MC 249-17, Pasadena, CA 91125, USA }
\begin{document}
\date{Accepted . Received ; in original form }

\pagerange{\pageref{firstpage}--\pageref{lastpage}} \pubyear{2002}
\maketitle
\label{firstpage}
\begin{abstract}
We report a dynamically evolving low ionization 
broad absorption line flow in the QSO SDSS J133356.02+001229.1 (at 
\zem $\sim$ 0.9197). These observations are part of our ongoing 
monitoring of low ionization broad absorption line (BAL) 
QSOs with the 2m telescope at IUCAA Girawali observatory (IGO). 
The broad \mgii absorption with an ejection velocity of  1.7$\times10^4$ \kms,
found in the Sloan Digital Sky Survey (SDSS) spectra, has 
disappeared completely in our IGO spectra. We found an emerging 
new component at an ejection velocity of  2.8$\times10^4$ \kms. 
During our monitoring period this component has shown strong evolution 
both in its velocity width and optical depth 
and nearly disappeared in our latest observations.
Acceleration of a low velocity component seen in SDSS spectrum to a
higher velocity is unlikely as the \mgii column densities are always 
observed to be higher for the new component. We argue that the
observed variations { may} not be {related}  to ionization 
changes and are consistent with absorption produced by multi-streaming
flow transiting across our line of sight.  We find a possible connection between
flux variation of the QSO and $N$(Mg~{\sc ii}) of the newly emerged component.
This could  mean the ejection being triggered by changes in the accretion
disk or dust reddening due to the outflowing gas. 
\end{abstract}


\section{Introduction}
It is widely believed that the kinetic energy output from quasars 
through the outflows are as important as their radiative output. 
They are believed to play an important role in regulating the growth
of supermassive blackholes \citep{Silk98,King03} and  
star formation in the host galaxies \citep[e.g.,][]{Bower06}. 
In addition, outflows influence the enrichment of the surrounding 
intergalactic medium. Therefore, it is important to understand
the origin and evolution of QSO outflows. Large scale outflows  
manifest itself in the form of broad absorption lines (BALs)
in QSO spectra. 
BAL QSOs comprise of up to 40\% of the total QSO population
\citep[][and references therein]{Dai08}. The correct explanation 
for the observed incidence of BAL QSOs  is still a subject of debate 
between the orientational 
\citep[see for example,][]{hines95, goodrich95, murray95,elvis00} 
and evolutionary 
models \citep[see for example,][]{hazard84,becker00}.

Line variability studies of  BAL QSOs are useful for understanding 
the physical conditions and dynamics of the gas close 
to the central engine. 
The time variability of
\civ  and Si~{\sc iv }absorption is reported in several cases
\citep[for example,][]{Srianand01,lundgren07,gibson08}. Such a variability
could be related either to variations in the ionization state or 
to the covering factor of the absorbing gas.
However, the most interesting cases are the ones where the flow emerges 
afresh or shows strong dynamical evolution (i.e variation in the 
absorption profile and signatures of acceleration).
There are {five} previous reports of emerging \civ\ BAL discovery in the 
literature [TEX 1726+344 \citep{ma02},  SDSS J105400.40$+$034801.2 \citep{hamann08}, {WPVS 007 \citep{Leighly09}}, Ton 34 \citep{Krongold10} and  PG0935+417 \citep{hidalgo11}].
These studies attribute the observed dynamical evolution to
multiple streaming wind moving across the line of sight.
Detecting an emerging Mg~{\sc ii} flow will be very interesting as
Mg~{\sc ii} BALs are considered to be a possible
manifestation of a QSO's efforts to expel a thick shroud of 
gas and dust \citep{Voit93}. 

\begin{table*}
\caption{Log of Observations}
\flushbottom
\begin{tabular}{ccccccccc}
\hline
Source&Instrument& Date&MJD&Exposure Time&$\lambda$ Coverage&Resolution&S/N$^a$\\
&&&&(min)&($\AA$)&(kms$^{-1}$)&\\
\hline
\hline
&SDSS&28-04-2000&51662&45x1&3800-9200&150&27\\
&SDSS&15-02-2001&51955&105x1&3800-9200&150&40\\
&IGO/IFOSC 7&03-04-2008&54559&40x2&3800-6840&300&11\\
&IGO/IFORS 1&26-02-2009&54888&45x3&3270-6160&360&21\\
&IGO/IFOSC 7&26-03-2009&54916&45x3&3800-6840&300&12\\
J1333+0012&IGO/IFOSC 7&19-04-2009&54940&45x2&3800-6840&300&5\\
&IGO/IFORS 1&22-01-2010&55218&45x1&3270-6160&360&19\\
&IGO/IFORS 1&25-01-2010&55221&45x1&3270-6160&360&16\\
&IGO/IFORS 1&14-03-2010&55269&45x3&3270-6160&360&16\\
&IGO/IFOSC 7&06-04-2011&55657&45x3&3800-6840&300&19\\
\hline
\end{tabular}
 \begin{flushleft}
  $^a$  calculated  over the wavelength range 5800-6200$\AA$ \\
    \end{flushleft}
 \label{log}
\end{table*}

Here,
we report the first discovery of an emerging Mg~{\sc ii} outflow in
a QSO SDSS J133356.02+001229.1(hereafter J1333+0012) at a redshift 
\zem $\sim$ 0.9197 \citep{Hewett01,Trump06}. 
This source is part of our sample of low ionization BAL QSOs that are being
spectroscopically monitored at IUCAA Girawali observatory (IGO).
Our spectra revealed the emergence of a new \mgii broad
absorption component that was not present in the Sloan Digital Sky Survey(SDSS) 
spectra obtained 7 years earlier. 
We have been monitoring this new Mg~{\sc ii} absorption component
for the past 4 years. 
In section 2, we provide details of the observations and data reduction. 
Results and discussions are presented in Sections 3 and 4 respectively.

\section{Observation and Data Reduction}
Our spectroscopic observations were carried out with 
IUCAA Faint Object Spectrograph (IFOSC) mounted on a 
2m telescope at IUCAA Girawali observatory (IGO).
The details of these observations together with that of 
the archival SDSS spectra  are given in Table~\ref{log}.  
We have used Grism \#1 and Grism \#7  of IFOSC 
in combination with a long slit having a width of
1.5 arcsec. This  combinations have the 
wavelength coverage of  3270$-$6160 $\AA$ and 
3800$-$6840$\AA$  respectively. Typically the observations 
were splitted in to exposures of 45 minutes. 
Spectra obtained on closely separated dates were combined 
for better signal to noise. Cleaning of the raw frames 
and 1D spectral extraction were carried out using the standards
procedures in IRAF{\footnote {IRAF is distributed by the National Optical Astronomy Observatories, 
which are operated by the Association of Universities for Research in Astronomy, Inc., 
under cooperative agreement with the National Science Foundation.}}. 
We opted for the variance-weighted extraction with ``doslit'' procedure.
Wavelength calibrations were done using standard helium neon lamp 
spectra and flux calibrations were done using a 
standard star spectra observed on the same night. 
Air-to-vacuum conversion was applied before adding the spectra.
 Individual spectra were combined using 1/$\sigma^2$ weighting in 
each pixel after scaling the overall individual spectra
 within a sliding window. The error spectrum  
was computed taking into account proper error propagation during the combining process.

\begin{figure}
 \centering
\begin{tabular}{c}
\psfig{figure=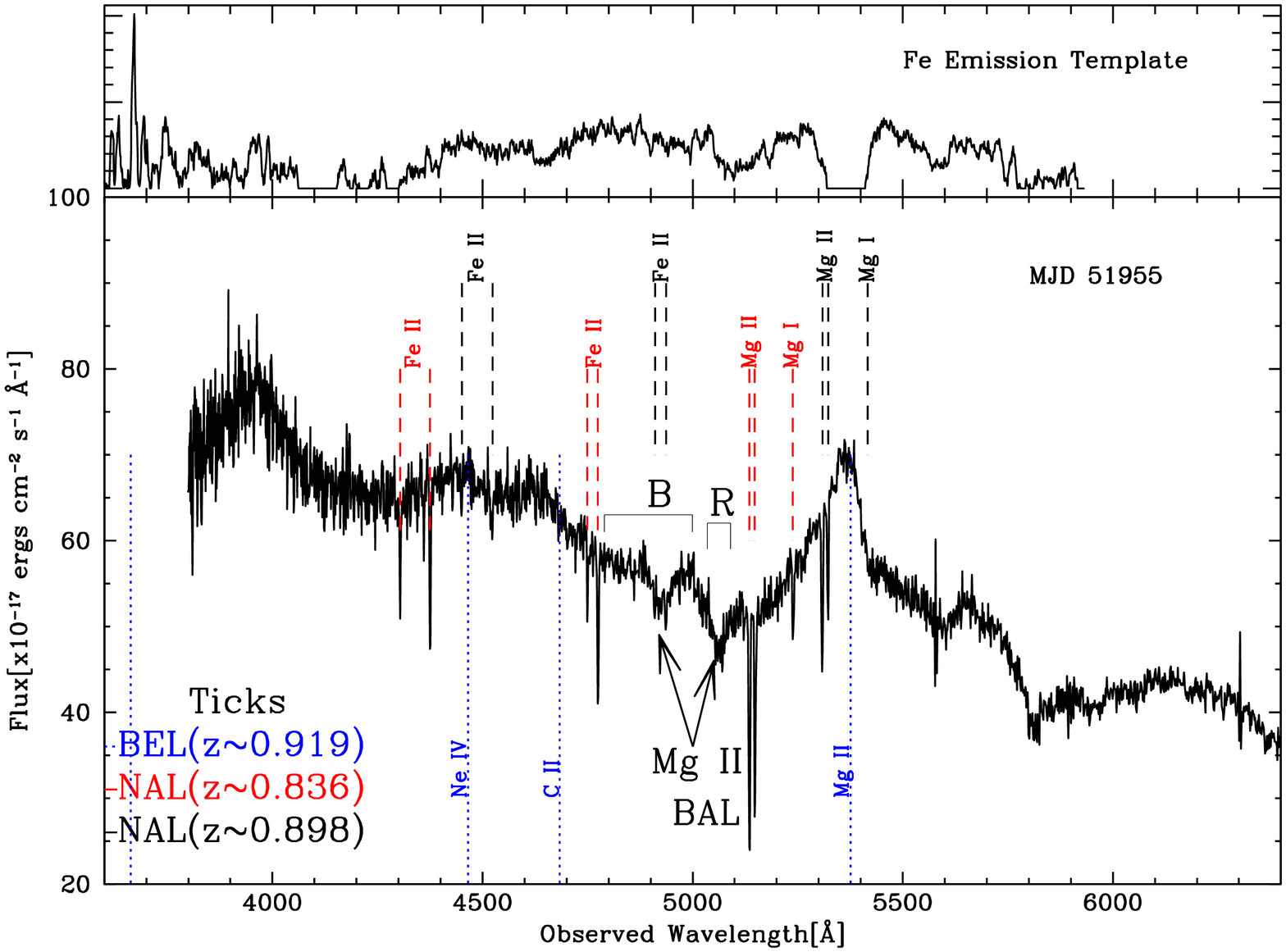,width=1.0\linewidth,height=0.7\linewidth,angle=270}
\end{tabular}
\caption{The SDSS spectrum of J1333+0012. The expected positions of 
broad emission lines are marked with blue dotted lines. The narrow 
absorption lines at \zabs $\sim$ 0.898 and 0.836 are shown by 
black and red dashed lines. Arrows show the red (marked as R)
and blue (marked as B) components of Mg~{\sc ii} BAL discussed in this
article. Fe emission template from \citet{Vestergaard01} is shown in 
the upper panel for comparison. }
 \label{fig1}
 \end{figure}

Fig.~\ref{fig1} shows the SDSS spectrum taken on 15-02-2001.
The expected positions of broad emission lines (at \zem=0.9197)  
are marked with blue dotted vertical lines.
Narrow  Mg~{\sc ii}, \mgi and  \feii absorption lines associated with the
intervening absorbers at \zabs$\sim$ 0.898 and \zabs$\sim$ 0.836  are
distinctly detected. In addition, two broad Mg~{\sc ii} absorption components
(marked in Fig.~\ref{fig1} as 'B' and 'R' to
denote the blue and red components respectively) are clearly detected. 
We do not detect any associated \feii lines and the \aliii lines are 
redshifted out of the coverage. 
A visual comparison of this spectrum with the Fe template of \citet{Vestergaard01} suggests that the continuum of J1333+0012 may have
considerable contribution from broad Fe emission lines.

\begin{figure*}
 \centering
\begin{tabular}{c c}
\psfig{figure=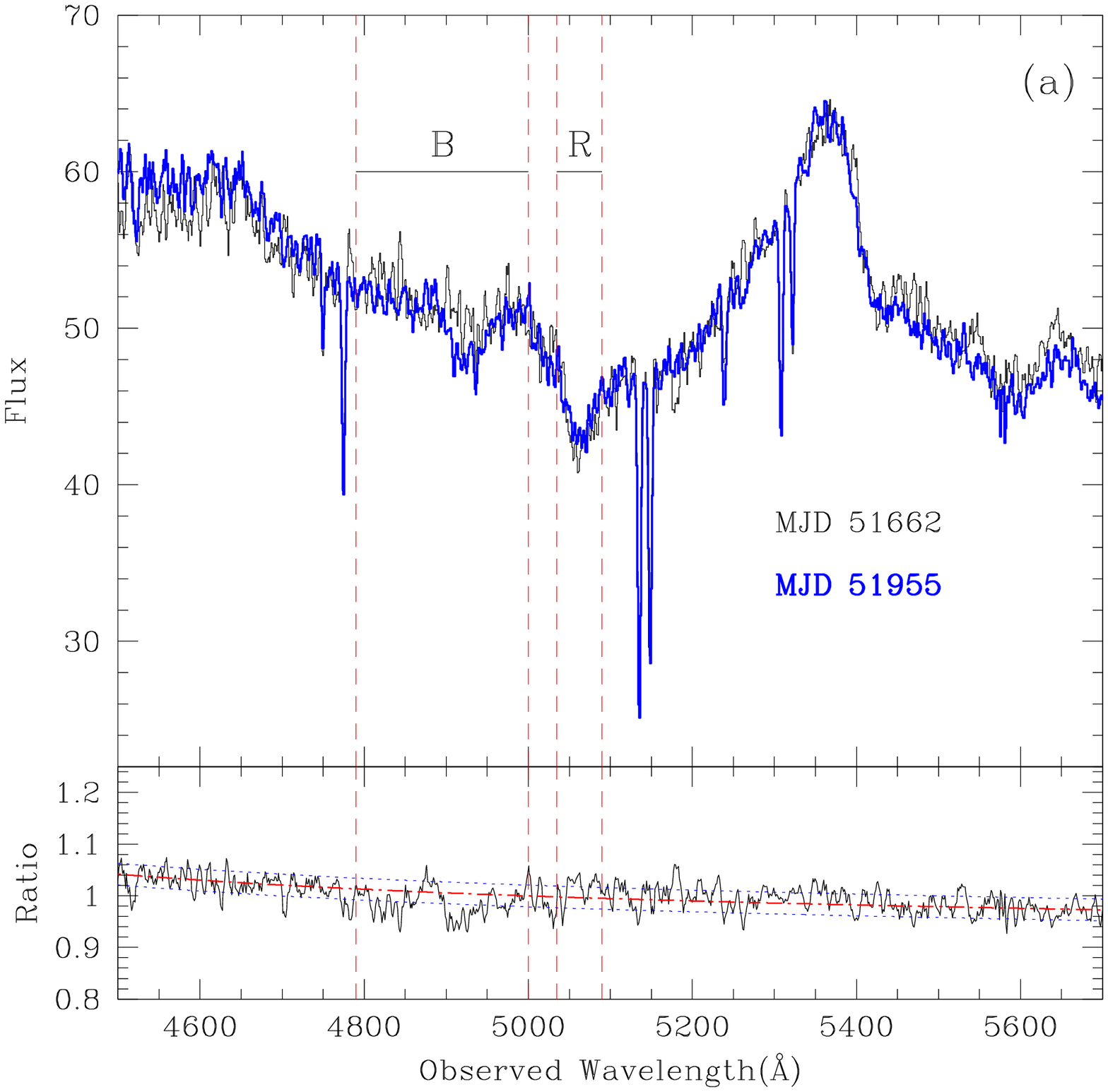,width=8cm,height=6.cm,bbllx=21bp,bblly=200bp,bburx=570bp,bbury=705bp,clip=yes}&
\psfig{figure=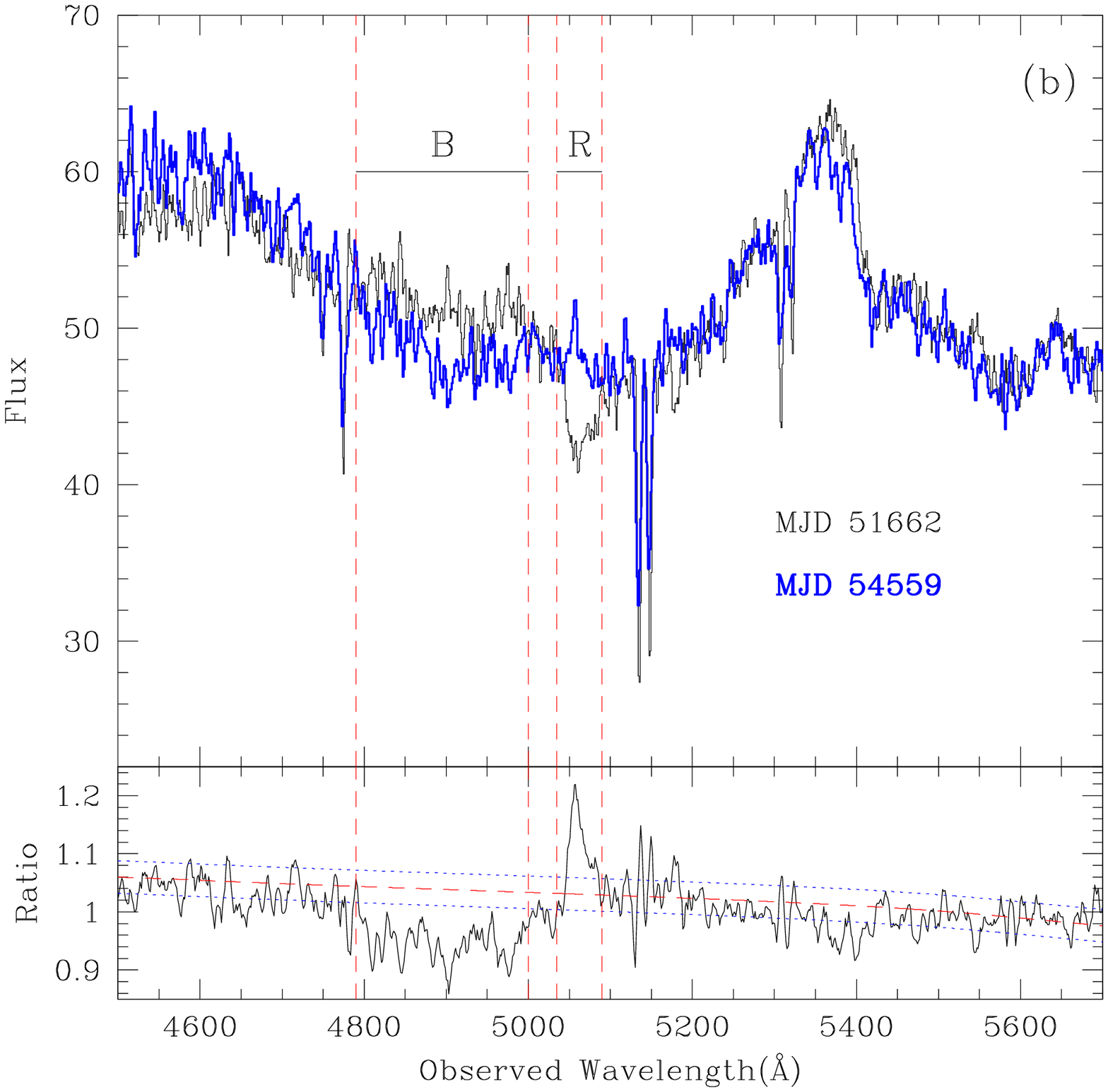,width=8cm,height=6.cm,bbllx=51bp,bblly=200bp,bburx=570bp,bbury=705bp,clip=yes}\\
\psfig{figure=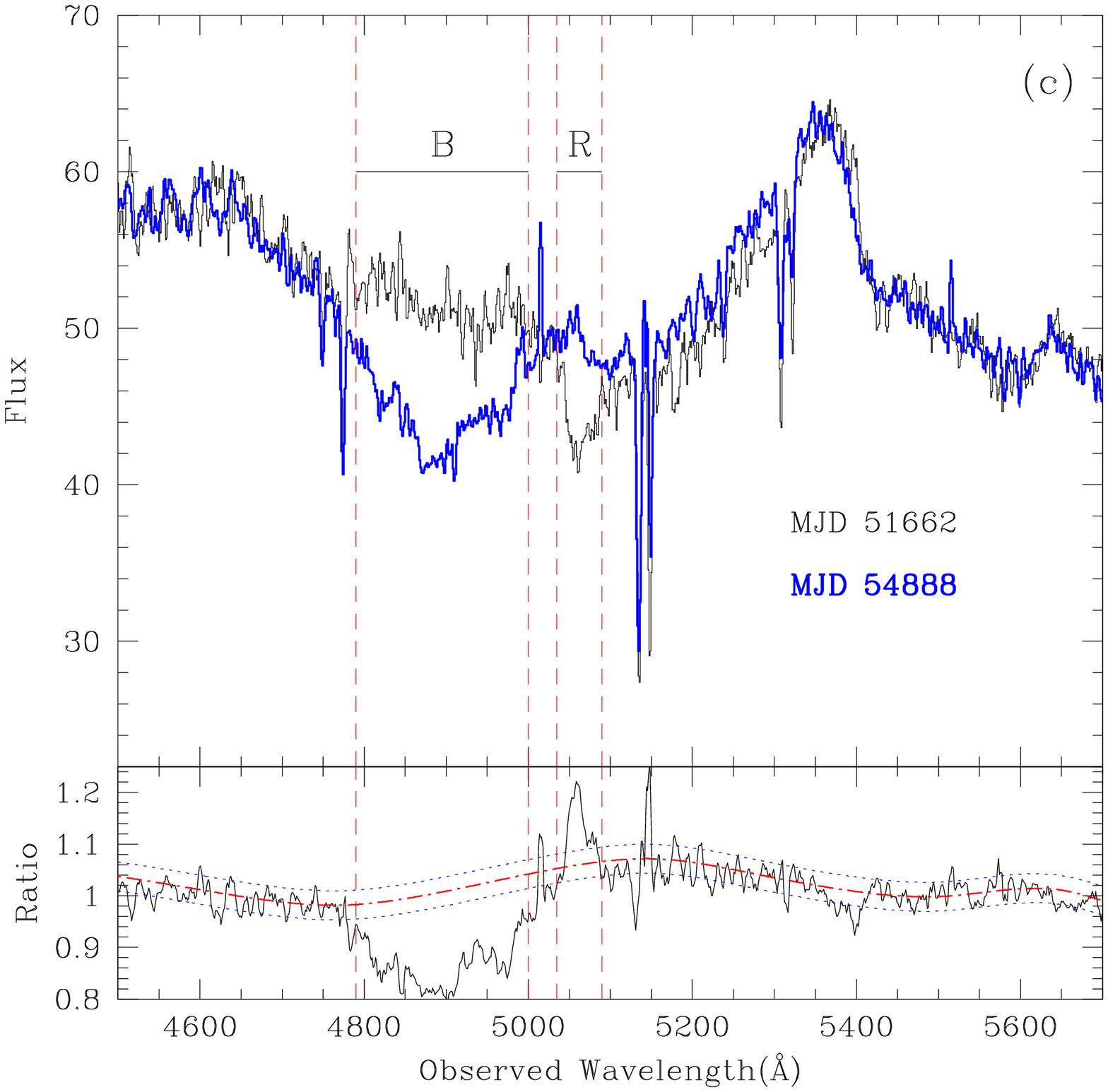,width=8cm,height=6.cm,bbllx=21bp,bblly=200bp,bburx=570bp,bbury=705bp,clip=yes}&
\psfig{figure=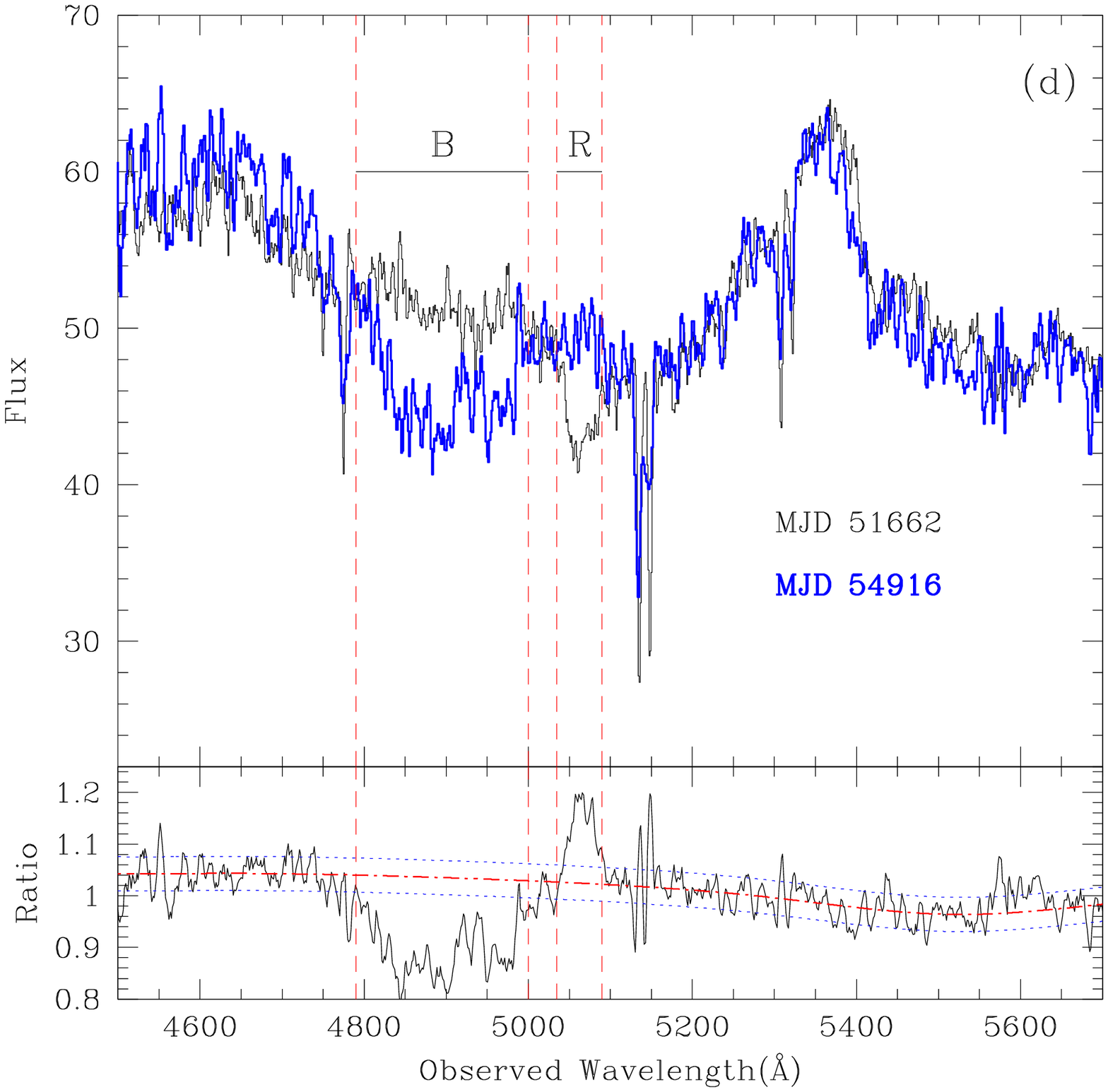,width=8cm,height=6.cm,bbllx=51bp,bblly=200bp,bburx=570bp,bbury=705bp,clip=yes}\\
\psfig{figure=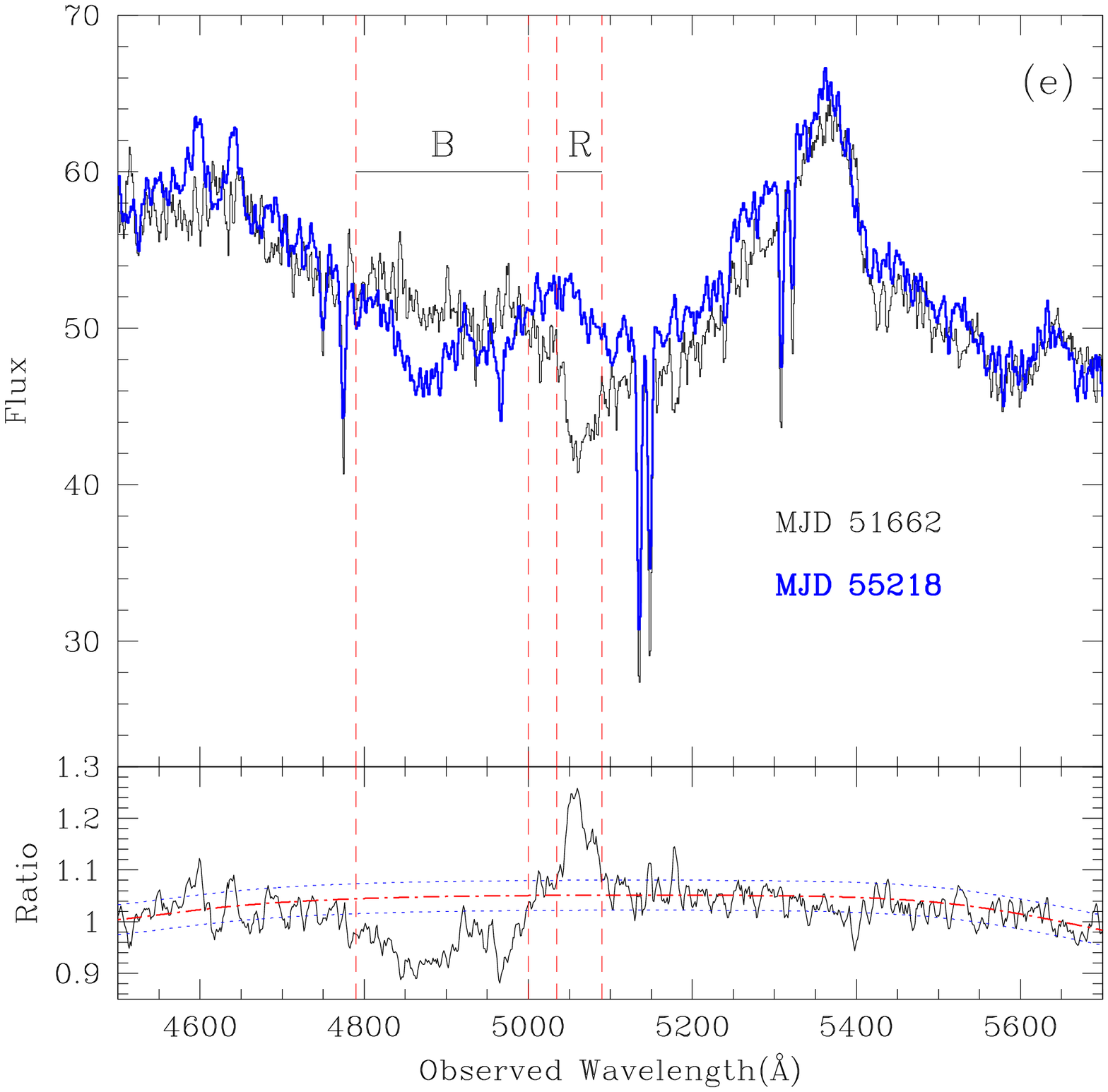,width=8cm,height=6.cm,bbllx=21bp,bblly=150bp,bburx=570bp,bbury=698bp,clip=yes}&
\psfig{figure=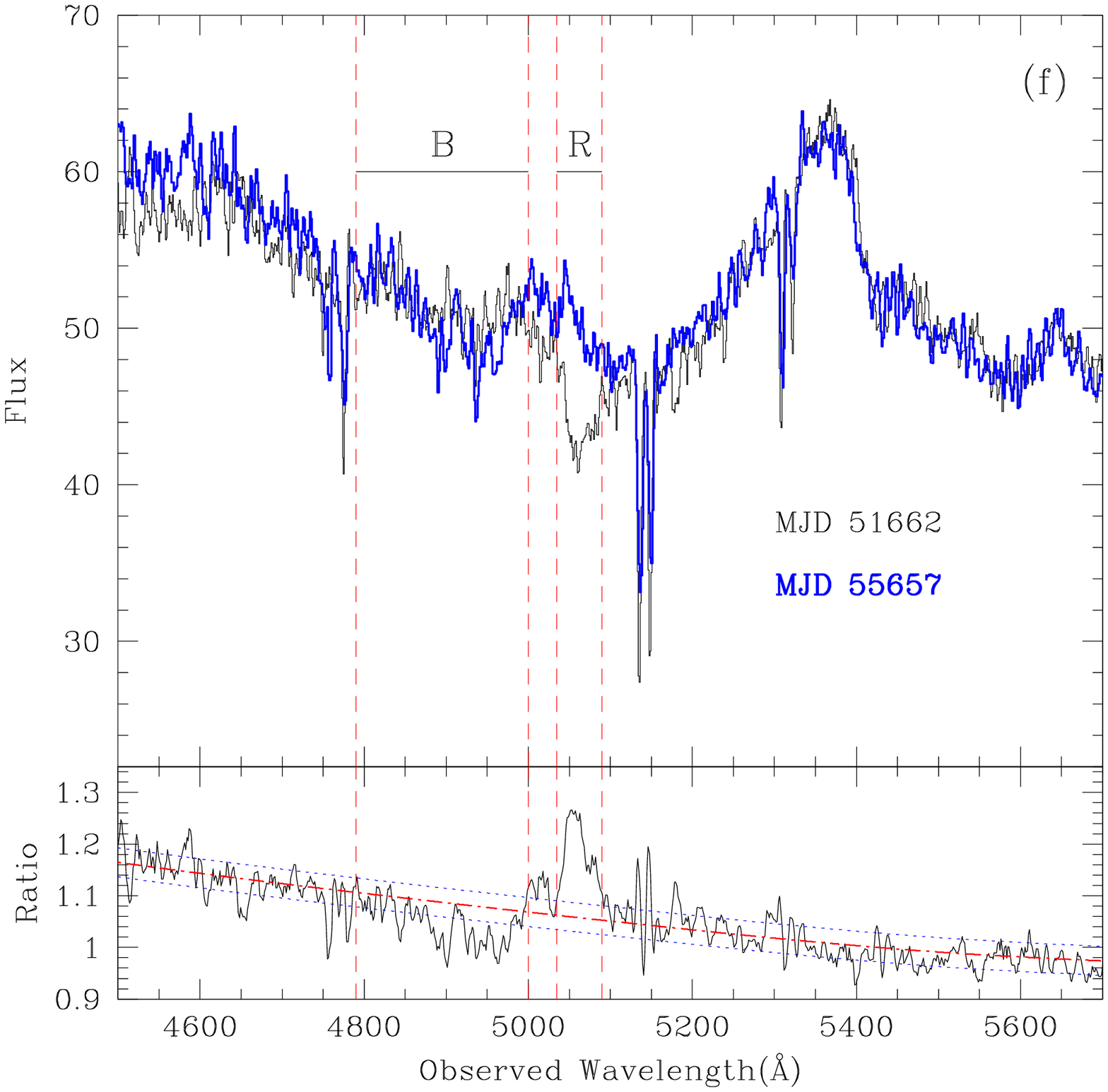,width=8cm,height=6.cm,bbllx=51bp,bblly=150bp,bburx=570bp,bbury=698bp,clip=yes}\\
\end{tabular}
\caption{The spectra of J1333+0012 obtained at different epochs are compared 
with the initial SDSS spectrum. The observed flux is scaled to match the SDSS flux. The vertical lines show the wavelength range of R and B components as in Fig.~\ref{fig1}.
The  ratio between the SDSS and the corresponding  IGO spectrum is plotted 
in the lower panel in each plot. A lower order polynomial fit 
to this curve and the associated error
are also shown. 
%
}
\label{fig2}
\end{figure*}

\section{Results}

\citet{Trump06} have fitted the spectrum of J1333+0012,
observed on 15-02-2001, using a 
template spectrum.  Their best fitted template resulted in a 
spectral index ($f_\nu\propto \nu^{\alpha_\nu}$) of $\alpha_\nu=1.64$ 
and E(B-V) = 0.295 for 
a SMC type dust. They have identified only one distinct 
BAL component at \zabs = 0.811 (corresponding to R component 
defined above). 
They assigned a Balnicity index and Absorption index of 36$\pm$0.42 and 459$\pm$0.75 respectively to this source. 
\begin{table}
\caption{Column Density Measurements}
\flushbottom
\setlength{\tabcolsep}{3.2pt}
\begin{tabular}{ccccccc}
\hline
MJD&\multicolumn{3}{c}{R Component}&\multicolumn{3}{c}{B Component}\\
\hline
   &log~$N$(\mgii)$^1$&$\Delta v^a$&z$_{abs}^b$&log~$N$(Mg~{\sc ii})$^1$\
&$\Delta v^a$&z$_{abs}^b$\\
&(cm$^{-2}$)&(kms$^{-1}$)&&(cm$^{-2}$)&(kms$^{-1}$)&\\
\hline
51662& 13.62$\pm$ 0.20& 2829 &0.811 & $\le$13.42$^*$ & 7000  & - \\
51955& 13.75$\pm$ 0.13& 3934 & 0.812 & 13.36$\pm$0.25& 6143  & 0.764	\\
54559& $\le$13.57$^*$ & 3000  & - & 14.26$\pm$0.39& 12303   &0.747 \\
54888& $\le$13.20$^*$ & 3000  & -& 14.50$\pm$0.10 & 12909  &0.746  \\ 
54916& $\le$13.42$^*$ & 3000  & -& 14.46$\pm$0.20 & 11900 & 0.745 \\
55218& $\le$13.09$^*$ & 3000  & -& 14.28$\pm$0.13& 11474 &0.754  \\
55657& $\le$13.35$^*$ & 3000  & -& 13.76$\pm$0.36& 10016&0.760   \\
\hline
\end{tabular}
 \begin{flushleft}
{ $^1$ { The error in $N$(\mgii) does not include the contribution 
from continuum placement uncertainties; $^a$ velocity range that contains 
90\% of the integrated optical depth \citep[see][]{Ledoux06a}; 
$^b$ optical depth weighted redshift};
$^*$ 3 $\sigma$ Upper Limit of $N$(\mgii). \\
} 
\end{flushleft}
 \label{tabcol}
\end{table}
Apart from this R component (that has consistent profile 
between two SDSS spectra
and the KECK spectra of \citet{Hewett01} observed on 30-04-2000) 
we find a new component at \zabs = 0.764 
(defined above as B) in the 2001 SDSS
data 
(not clearly detected in the previous epoch SDSS spectrum) 
having a velocity width of  $\sim$6000 \kms.  

In Fig.~\ref{fig2}, the IGO spectra obtained at different 
epochs are compared with a reference SDSS spectrum obtained in year 2000 
(i.e MJD 51662). 
{For displaying purpose
we have scaled the fluxes to match the flux of the reference spectrum.
In the bottom half of each panel we show the ratio of spectrum
obtained at a given epoch to that of the 
reference spectrum.
This curve is smooth apart from the wavelength range of R and B components.
We normalise this curve with a smooth lower order polynomial (shown as
red dotted curve in Fig.~\ref{fig2}) to remove any
large wavelength scale continuum flux differences between the two spectra.}
%

In our IGO 2008 (MJD 54559) spectrum (see panel b in Fig~\ref{fig2}), 
the R component ( that was seen with near identical
optical depth in two SDSS spectra) has completely disappeared.  
The R component has never reappeared again in any of our subsequent
IGO spectra (see Fig~\ref{fig2}).
At the same time, the B component is seen to have varied in depth 
and widened in velocity in the 2008 (MJD 54559) data. 
In 2009 (MJD 54888), 
the B component further deepened in the absorption troughs. But, no changes in the velocity structure are seen (see panels c and d in Fig.~\ref{fig2}). 
It  diminished in strength in 2010 and  has almost vanished in 2011 (see
panels e and f in Fig.~\ref{fig2}).  

We calculated the Mg~{\sc ii} column density 
for both R and B components using apparent optical depth method 
\citep{Savage91} assuming the gas to be optically thin and covering 
the background source completely. { In case of the component B,
we have used the normalised ratio spectra discussed in Fig.~\ref{fig2}
to estimate $N$(\mgii). For the component R that is relatively narrow
we use the observed flux in the neighborhood of the absorption line
for continuum normalization.}
%
The measured \mgii\ column densities
(or 3$\sigma$ upper limits), velocity width ($\Delta v$) and \zabs\
 for both the components are summarized in Table~\ref{tabcol}.

\begin{figure}
\centering
\psfig{figure=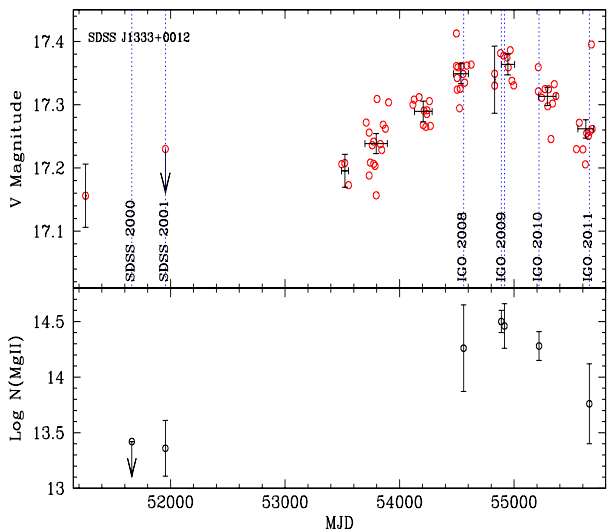,width=1.0\linewidth,height=0.9\linewidth,angle=270}
\caption{{\it Top panel:} Light curve (in Johnson's V magnitude) 
of J1333+0012 from CRTS and SDSS (first two points) measurements. 
Black points are the average of closely spaced observations 
and the associated error is obtained from errors in individual measurements.  The x-error bars corresponds to the time period over which the individual measurements are averaged.
The epochs of spectroscopic observations are marked as blue dotted vertical lines. {\it Bottom panel:} the measured $N$(\mgii) of the B component vs MJD.}
\label{fig3}
\end{figure}

%
{ In Fig.~\ref{fig3} we plot the photometric light curves of
J1333+0012 obtained in the Catalina Real-Time Transient Survey
\citep[CRTS,][]{Drake09}. These observations were carried out
%
%
with an unfiltered set up and
were converted to V magnitude 
using 10 to 100 well selected G-type dwarf calibration stars found in
each frame and the method described in \citet{Henden00}. 
The zero point uncertainty
is found to be $\le$ 0.08 mag.}
The two left most points in the light curve corresponds to the 
transformed SDSS photometric and fiber magnitudes respectively. 
{ We use the transformation equation given in \citep{Jester05} 
for $z\le2.1$ QSOs for this.}
Fiber magnitude should be considered as an upper limit on 
the actual magnitude.  
%
{ The QSO has 
%
dimmed between the early epoch CRTS observations and IGO 2008 observations, 
reached a minimum in 2009 and then brightened thereafter.  
The maximum seen amplitude of the variation is $\sim$0.2 mag.}
%
Very interestingly, a similar but opposite variation is seen in the 
$N$(\mgii) of the B component during this period. i.e., when 
the quasar flux was minimum, the measured $N$(Mg~{\sc ii}) is maximum.
%

\section{Discussion}

The main result of our monitoring of J1333+0012 is that the 
R component of \mgii absorption has disappeared completely 
after the year 2001 and the B component emerged, became stronger
and wider and subsequently disappeared between the years 2001 and 2011. 
There is also tentative evidence for a possible connection between the 
QSO flux and $N$(Mg~{\sc ii}) measured for the B component.
Now we discuss various scenarios to explain these observations.

It is  very tempting to say that the R component got accelerated 
to the B component and later vanished from our line of sight. 
From Table~\ref{tabcol}, it is clear that the value of $N$(Mg~{\sc ii}) and
$\Delta v$ of component B in later epochs are much higher 
than what is seen in the R component in earlier epochs. If we use 
thin shell approximation, then mass conservation suggests that
$r^2 N$(H) should be conserved, where $r$ and $N$(H) are the
distance of the shell from the QSO and total hydrogen column 
density respectively. We expect $r$ to increase with time so
$N$(H) (and hence $N$(Mg~{\sc ii}) should decrease with time). 
The observed $N$(Mg~{\sc ii}) values do not not support this scenario.

One could argue that there were always two fluid components and what we 
see is due to the variations in the ionization parameter. 
Above we have seen that the QSO has varied by $\sim 0.2$
mag. 
However the required change in $N$(Mg~{\sc ii}) is 0.6 to 1 dex.
It is clear from Fig.~2 of \citet{Hamann97} that required 
change in $N$(Mg~{\sc ii}) can not be produced by 0.2 dex 
change in ionization parameter. In addition the change in the 
Mg~{\sc ii} optical depth of R and B components are in the opposite 
directions. { This scenario can be consistent if the
variation in UV continuum is much larger than the 0.2 mag seen
in the V band and the initial ionization parameters of the two
components are very different. We do not favor this scenario
based on the lack of large colour variations between our IFOSC spectra
and the near consistency of \mgii\ emission line.}


In all the previous cases of emerging flows
\citep{ma02,hamann08,Krongold10,hidalgo11,Hall11} 
the most favoured explanation is the transverse motion of the absorbing gas. 
In particular the evolution of the B component can be explained as the 
absorbing gas transiting through our line of sight as  the component 
has emerged, evolved and disappeared. To 
consider this possibility we derive some basic parameters for J1333+0012.

We obtained the B band magnitude of this QSO using the measured u and g 
magnitudes \citet[see][]{Jester05}. We use this and the prescription 
of \citet{marconi04}, to get the bolometric luminosity,  
L$_{bol}$ = 9.39 $\times$ 10$^{46}$ ergs s$^{-1}$. The corresponding 
mass accretion rate, $\dot{M_{in}}$ is,
\begin{equation}
\centering
 \dot{M_{in}} = \frac{L_{bol}}{\epsilon c^2} = 16.9 M_{\odot}/yr 
 \end{equation}
 \label{eqn4}
where $\epsilon$, the mass to energy conversion efficiency is taken as 0.1.  
The blackhole mass derived from Eddington accretion is 8$\times 10^8$ M$_\odot$.
Following \citet{Hall11}, { we find the diameter of the disc within 
which 90\% of the 2700-\AA~  continuum is emitted
D$_{2700} \sim 10^{16}$ cm.}
 
The variability of the R component has occurred over a period 3994 days
in the observers frame or 2081 days in the rest frame of the QSO. 
We can use this as an upper limit on the transit time of the absorbing
gas across the continuum emitting region. If we assume the { projected
transverse size} of the absorbing gas is much smaller than the emitting region
(assumed to be a face-on disk), 
then we can estimate the transverse velocity $v_\perp\ge 550$ \kms. 
%
We find the radial velocity up to $v_r$ = 2.8$\times 10^4$ \kms and the
radial displacement of the gas during the transit is $\sim 0.1 pc$.
Even if the { projected transverse}  size of the gas is 10 times the continuum emitting region,
the expected transverse velocity will be much lower than the measured
radial velocity. 

The main question is,
is the emergence of new component in any way related to the structural
changes in the accretion disk?. The possible connection between
QSO dimming and $N$(Mg~{\sc ii}) is interesting. It could either
mean that the ejection being triggered by some events in the accretion disk
that caused reduction in the accretion efficiency or extinction by
dust in the transiting gas. 

\citet{Trump06} identified this system as BAL based on the presence of
Mg~{\sc ii} absorption in the R component. The latest spectrum of
J1333+0012 taken with IGO is nearly devoid of any broad \mgii\ absorption. 
We do not detect Mg~{\sc i}, \feii\ and Ca~{\sc ii} absorption associated 
with either of the broad \mgii\ components. It will be interesting to 
study the properties of other strong high ionization lines using
UV spectroscopy. Continuous optical monitoring of this source
to see recurrence of new outflow episode will be useful for 
modeling this interesting source.

\section{Conclusions}
In this paper, we report a dynamically evolving \mgii broad absorption 
line outflow in SDSS J133356.02+001229.1. The R component detected in 
the SDSS spectra taken in 2000 and 2001 was never seen in the IGO spectra taken
in the latter epochs. The blue component B emerged in 2001. 
It continued to increase its optical depth till 2009 and started to 
diminish thereafter. This component has almost vanished in the most 
recent spectra in 2011. 

We provide arguments for not favoring the acceleration of a single 
cloud to a higher 
velocity and changes in the ionization conditions causing the 
observed spectral variability. The observed variability can be best explained 
by multiple streaming gas moving across our line of sight. 
An outflowing gas component
with transverse velocity one tenth of the observed radial velocity
and flow extent up to 10 times the size of the accretion disk can 
easily explain these observations.

We find the QSO has dimmed by about $~$0.2 mag (in V-band) 
during the monitoring
period. This may be the manifestation of the emerged outflow being
triggered by the structural changes in the accretion disk or due
to reddening by dust in the absorbing gas.


\section{acknowledgements}
{We wish to acknowledge the IUCAA/IGO staff for their support during our observations.  In particular we thank Dr. Vijay Mohan for continued support. 
MV gratefully acknowledges University
Grants Commission, India, for support through RFSMS Scheme
and IUCAA for hospitality, where most of this work was done.
RS gratefully acknowledge the support from the Indo-French Center for
the Promotion of Advanced Research (Centre Franco-Indien pour la promition de la recherche avanc\'ee) under project N. 4304-2.
}



\def\aj{AJ}%
\def\actaa{Acta Astron.}%
\def\araa{ARA\&A}%
\def\apj{ApJ}%
\def\apjl{ApJ}%
\def\apjs{ApJS}%
\def\ao{Appl.~Opt.}%
\def\apss{Ap\&SS}%
\def\aap{A\&A}%
\def\aapr{A\&A~Rev.}%
\def\aaps{A\&AS}%
\def\azh{AZh}%
\def\baas{BAAS}%
\def\bac{Bull. astr. Inst. Czechosl.}%
\def\caa{Chinese Astron. Astrophys.}%
\def\cjaa{Chinese J. Astron. Astrophys.}%
\def\icarus{Icarus}%
\def\jcap{J. Cosmology Astropart. Phys.}%
\def\jrasc{JRASC}%
\def\mnras{MNRAS}%
\def\memras{MmRAS}%
\def\na{New A}%
\def\nar{New A Rev.}%
\def\pasa{PASA}%
\def\pra{Phys.~Rev.~A}%
\def\prb{Phys.~Rev.~B}%
\def\prc{Phys.~Rev.~C}%
\def\prd{Phys.~Rev.~D}%
\def\pre{Phys.~Rev.~E}%
\def\prl{Phys.~Rev.~Lett.}%
\def\pasp{PASP}%
\def\pasj{PASJ}%
\def\qjras{QJRAS}%
\def\rmxaa{Rev. Mexicana Astron. Astrofis.}%
\def\skytel{S\&T}%
\def\solphys{Sol.~Phys.}%
\def\sovast{Soviet~Ast.}%
\def\ssr{Space~Sci.~Rev.}%
\def\zap{ZAp}%
\def\nat{Nature}%
\def\iaucirc{IAU~Circ.}%
\def\aplett{Astrophys.~Lett.}%
\def\apspr{Astrophys.~Space~Phys.~Res.}%
\def\bain{Bull.~Astron.~Inst.~Netherlands}%
\def\fcp{Fund.~Cosmic~Phys.}%
\def\gca{Geochim.~Cosmochim.~Acta}%
\def\grl{Geophys.~Res.~Lett.}%
\def\jcp{J.~Chem.~Phys.}%
\def\jgr{J.~Geophys.~Res.}%
\def\jqsrt{J.~Quant.~Spec.~Radiat.~Transf.}%
\def\memsai{Mem.~Soc.~Astron.~Italiana}%
\def\nphysa{Nucl.~Phys.~A}%
\def\physrep{Phys.~Rep.}%
\def\physscr{Phys.~Scr}%
\def\planss{Planet.~Space~Sci.}%
\def\procspie{Proc.~SPIE}%
\let\astap=\aap
\let\apjlett=\apjl
\let\apjsupp=\apjs
\let\applopt=\ao
\bibliographystyle{mn}

\bibliography{1333}
\label{lastpage}

\end{document}